\documentclass[%
 reprint,
 amsmath,amssymb,
 aps, prl
]{revtex4-2}

\usepackage{graphicx}% Include figure files
\usepackage{dcolumn}% Align table columns on decimal point
\usepackage{bm}% bold math
\usepackage{amsmath}
\usepackage{physics}
\usepackage[dvipsnames]{xcolor}
\begin{document}

\preprint{APS/123-QED}

\title{Cavity-mediated localization and collective electron correlation phases}% Force line breaks with \\
\author{Dominik Sidler}
 %\homepage{https://www.zhaw.ch/en/engineering/institutes-centres/icp-institute-of-computational-physics/fundamental-physics-industrial-applications/quantum-light-matter-engineering}
 \affiliation{Zurich University of Applied Sciences, School of Engineering, 
 Technikumstrasse 71, Winterthur 8400, Switzerland}
\affiliation{%
 Max Planck Institute for the Structure and Dynamics of Matter and Center for Free-Electron Laser Science, Luruper Chaussee 149, Hamburg 22761, Germany
}%Lines break automatically or can be forced with \\
 \email{dominik.sidler@zhaw.ch}
 
%\author{Francesco Greco}
% \affiliation{Max Planck Institute for the Structure and Dynamics of Matter and Center for Free-Electron Laser Science, Luruper Chaussee 149, Hamburg 22761, Germany}%Lines break automatically or can be forced with \\
%\affiliation{Max Planck Institute for the Structure and Dynamics of Matter and Center for Free-Electron Laser Science, Luruper Chaussee 149, Hamburg 22761, Germany}
%\affiliation{The Hamburg Center for Ultrafast Imaging, Luruper Chaussee 149, 22761 Hamburg, Germany}

%


\author{Francesco Greco}%
\affiliation{Max Planck Institute for the Structure and Dynamics of Matter and Center for Free-Electron Laser Science, Luruper Chaussee 149, Hamburg 22761, Germany}
\affiliation{The Hamburg Center for Ultrafast Imaging, Luruper Chaussee 149, 22761 Hamburg, Germany}

\author{Michael Ruggenthaler}%
\affiliation{Max Planck Institute for the Structure and Dynamics of Matter and Center for Free-Electron Laser Science, Luruper Chaussee 149, Hamburg 22761, Germany}
\affiliation{The Hamburg Center for Ultrafast Imaging, Luruper Chaussee 149, 22761 Hamburg, Germany}

\author{Angel Rubio}
\affiliation{Max Planck Institute for the Structure and Dynamics of Matter and Center for Free-Electron Laser Science, Luruper Chaussee 149, Hamburg 22761, Germany}
\affiliation{The Hamburg Center for Ultrafast Imaging, Luruper Chaussee 149, 22761 Hamburg, Germany}
\affiliation{Center for Computational Quantum Physics (CCQ), Initiative for Computational Catalysis (ICC), The Flatiron Institute, 162 Fifth avenue, New York, NewYork 10010, United States of America}

\collaboration{UnMySt Collaboration}%\noaffiliation

\date{\today}% It is always \today, today,

\begin{abstract}
 Collective strong coupling of molecular ensembles to optical cavities opens a route to modifying matter through genuinely collective electronic correlations. Yet even in the absence of a cavity, Coulomb correlations are notoriously difficult to describe, and cavity coupling adds transverse correlation channels extending over the entire molecular ensemble. Here we show that this seemingly intractable problem admits a controlled description by spin glass theory, i.e., by the analytically solvable spherical Sherrington-Kirkpatrick model. Our results predict two collective correlation phases, a paracorrelated phase and a spin-glass correlation phase, beyond the conventional uncorrelated molecular regime. These phases reveal an entropy-driven localization-delocalization mechanism that transfers molecular electronic states into collectively degenerate cavity-dressed states. Analytic calculation suggest that this collective correlated state can either remain insulating or even turn metallic if the entropic occupations favor a fractional filling.  
 Our work reveals  cavity-mediated electron correlations as a microscopic mechanism for emergent phases in strongly coupled molecular ensembles.
 %Collective strong coupling of molecular ensembles in optical cavities still poses numerous open theoretical questions, among which   cavity-modified electron correlations appear to be critical for many of the observed phenomena. However, accurately treating Coulomb  correlations is already a notoriously difficult problem even outside a cavity. Inside a cavity, additional transversal collective electron correlation effects span across the molecular ensemble and thus seem completely inaccessible. Neverthelsee, building on a recently proposed mapping of the collective transversal electron correlations to the analytically solvable spherical Sherrington-Kirkpatrick  spin glass model, we identify two collectively correlated phases, the \textit{paracorrelated} and the \textit{spin glass correlation} phase, in addition to the familiar uncorrelated phase. Ultimately, these collective correlations point to an entropic (de)localization mechanism that promotes molecular orbitals into collectively correlated states.   
%\begin{description}
%\end{description}
\end{abstract}

\maketitle

%%% -----------------------------------------------------------------
%%%  INTRODUCTION
%%% -----------------------------------------------------------------

Over the past decades, seminal experiments have demonstrated that collective strong coupling of matter in optical cavities can be utilized to alter and design material and molecular properties.~\cite{ebbesen_hybrid_2016,ruggenthaler_quantum-electrodynamical_2018,sidler_perspective_2022,fregoni_theoretical_2022, ebbesen_introduction_2023, ruggenthaler_understanding_2023, bhuyan_rise_2023, hirai_molecular_2023,simpkins_control_2023,mandal_theoretical_2023,xiang_molecular_2024}  Despite considerable theoretical advances, the relevant underlying physical mechanisms remain largely unknown. One of fundamental question is how collective strong coupling can be sufficiently localized such that they can induce significant chemical (local) changes as observed in experiments. In the following work, we focus on the collective strong coupling of molecular ensembles.  Cavity-modified electron correlations appear to be critical for many of the observed changes. However, accurately treating Coulomb  correlations is already a notoriously difficult problem to solve outside of a cavity. Inside a cavity, transversal collective electron correlation effects spanning  the molecular ensemble seem completely inaccessible. However, thanks to a recently proposed mapping of the collective transversal electron correlations to the analytically solvable spherical Sherrington-Kirkpatrick model of a spin glass, we can subsequently identify and investigate different phases of collective electron correlations. Notably, the design of spin glass conditions in optical cavities by structuring the photonic environment has been extensively discussed in Refs. \cite{gopalakrishnan_frustration_2011,gopalakrishnan_exploring_2012,guo_emergent_2019,guo_sign-changing_2019,marsh_entanglement_2024}. In our setup, however, the glassy picture emerges from the complex (random) molecular structure and the Pauli principle of the electrons.\cite{sidler_collectively-modified_2026}  To investigate collective transverse electron correlations, we extend recently proposed theoretical framework from collective configuration interaction singles\cite{szabo_modern_2012,helgaker_molecular_2000} (CIS) to a multireference / approximate Full CI picture, restricted\cite{olsen_determinant_1988} to a collectively degenerate subspace. Analytic considerations not only reveal novel, cavity-induced, phases of matter, but also suggest different mechanisms  (insulating vs. conducting) to enter such phases.

%%% -----------------------------------------------------------------
%%%  MODEL: EFFECTIVE SINGLE-ELECTRON PROBLEM
%%% -----------------------------------------------------------------

Within the cavity-Born-Oppenheimer partitioning, the electronic Hamiltonian is given as,\cite{ruggenthaler_understanding_2023}
\begin{widetext}
\begin{align}
%H &=    \sum_{i=1}^{2N}\bigg\{\frac{\hat{\vec{p}}_i^2}{2m} +V_{\rm ext}(\hat{\vec{r}}_i)+\frac{1}{2}\sum_{j=1,j\neq i}^{2 N}\frac{e^2}{4\pi \epsilon_0|\hat{\vec{r}}_i - \hat{\vec{r}}_j|}\bigg\}\nonumber\\
%&
%
%-\vec{\lambda}\sum_{i=1}^{2N}e\hat{\vec{r}}_i \bigg(\vec{\lambda}\sum_{i=1}^{N_{\rm nuk}}Z_n\vec{R}_i-\omega_\beta  %q_\beta\bigg)+\frac{\big(\vec{\lambda}\sum_{i=1}^{2N}e\hat{\vec{r}}_i\big)^2}{2} 
\hat{H}^{e} &=    \sum_{i=1}^{2N}\bigg\{\frac{\hat{\vec{p}}_i^2}{2m} +V_{\rm Coul}(\hat{\vec{r}}_i,\boldsymbol{R})-\vec{\lambda} \cdot e\hat{\vec{r}}_i \bigg(\vec{\lambda}\cdot \sum_{j=1}^{N_{\rm nuk}}Z_je\vec{R}_j-\omega_\beta  q_\beta\bigg)\bigg\}\nonumber\\
&
+\underbrace{\frac{1}{2}\sum_{i=1}^{2N}\sum_{j=1,j\neq i}^{2 N}\frac{e^2}{4\pi \epsilon_0|\hat{\vec{r}}_i - \hat{\vec{r}}_j|}}_{\hat{W}_\parallel}+\underbrace{\frac{\big(\vec{\lambda} \cdot \sum_{i=1}^{2N}e\hat{\vec{r}}_i\big)^2}{2} }_{\hat{W}_\perp}.
\label{eq:H}
\end{align}
\end{widetext}
The first line contains solely one-electron terms, with kinetic energy, Coulomb attraction of the nuclei $V_{\rm Coul}$ and the linear coupling to a single effective cavity mode $q_\beta$ of frequency $\omega_\beta$, linear polarization $\vec{\lambda}/|\lambda|$ and light-matter coupling parameter $\lambda=|\vec{\lambda}|$. The number of electrons is $2N$ and the positive nuclei enter with $\vec{R}_j$ and charge $Z_j e$. The second line, contains longitudinal (Coulomb) two-electron operators $\hat{W}_\parallel$ and the transversal dipole self-energy $\hat{W}_\perp$.
%

%We start by expressing the total energy of the \textit{ensemble} of interacting molecules, i.e., we consider the wave function of all the molecules in the cavity, in terms of the one-body reduced density matrix (1-RDM) $\gamma$ and the two-body reduced density matrix (2-RDM) $\Gamma$. Using the natural orbitals $\phi_p$~ we find the full energy by minimizing
%\begin{eqnarray}
%    E &=& \sum_{pq} h_{pq} \gamma_{pq} + \frac{1}{2}\sum_{pqrs}\bra{pq}rs\rangle \Gamma_{pqrs},\\
%    &=& \underbrace{\sum_{pq} h_{pq} \gamma_{pq} +\frac{1}{2}\sum_{pq} n_p n_q (2J_{pq}-K_{pq})}_{=E_{\rm EHF}} \nonumber\\
%    &&+ \underbrace{\frac{1}{2}\sum_{pqrs}\bra{pq}rs\rangle \Lambda_{pqrs}}_{=E_{\rm corr}}\label{eq:ensemble}
%\end{eqnarray}
%with $\gamma_{pq}=Tr[\hat{\Gamma}a_q^\dagger a_p]$, 2-RDM $\Gamma_{pqrs}=Tr[\hat{\Gamma}a_q^\dagger a^\dagger_p a_s a_r]$ and density-matrix operator $\hat{\Gamma}=\sum_k w_k \ket{\Psi_k}\bra{\Psi_k}$ with normalized ensemble weights $\sum_k w_k =1$. If the ground state of the ensemble is non-degenerate, we would find a pure state, i.e., $w_0 =1$ with a correlated ground-state wave function $\Psi_0$. Here we have used that we can re-express the full 2-RDM in terms of a Hartree-Fock-like term and corrections $\Gamma_{pqrs}=\gamma_{pr}\gamma_{qs}-\gamma_{ps}\gamma_{qr}+\Lambda_{pqrs}$, and the diagonal of the 1-RDM is given as $\gamma=\sum_p n_p \ket{\phi_p}\bra{\phi_p}$ for all the natural orbitals,\cite{mazziotti_reduceddensitymatrix_2007} with fractional occupations $0\leq n_p\leq 2$ that sum up to the total umber of electrons $\sum_p n_p=2N$.

\begin{figure}
    \centering
    \includegraphics[width=1\linewidth]{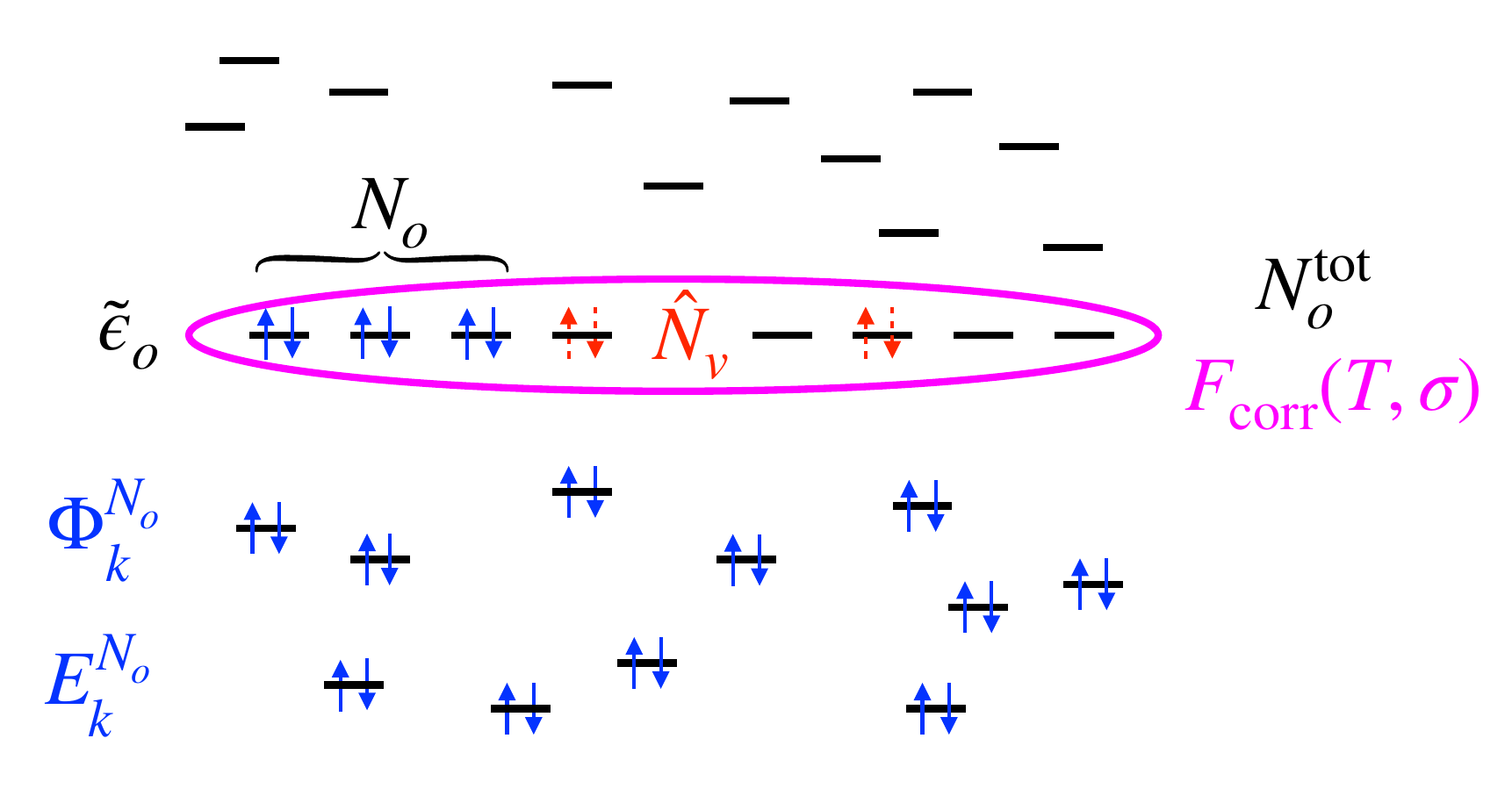}
    \caption{ Energies $\tilde{\epsilon}_i$ are determined  Hartree-Fock calculations.  A single reference state $\Phi_k^{N_o}$ with ensemble Hartree-Fock energy $E_k^{N_o}$ and $N_o$ occupied orbitals at degenerate HF energy $\tilde{\epsilon}_o$  is chosen and visualized in blue. Focusing on the $N_o^{\rm tot}$-fold degenerate subspace, we can start from the reference  state $|\Phi_k^{N_o}\rangle$ and investigate the transversal correlation (free)-energy $F_{\rm corr}$ using the Configuration Interaction (CI) method, i.e., by populating all combinations of $\hat{N}_v$ virtual orbitals (red).}
    \label{fig:occupation}
\end{figure}

Next, we aim to investigate transversal electron correlation effects that emerge from $\hat{W}_\perp$  under collective strong coupling and may persist in the thermodynamic limit. For this purpose, we focus on  molecular ensemble  degeneracies. 
The Hamiltonian given in Eq.~\eqref{eq:H} describes not only a single molecule but a full ensemble of molecules. Thus there are many rearrangements of nuclei/ions and symmetry transformations that leave the Hamiltonian (at least approximately) invariant. Consequently, if we find one Hartree-Fock Slater determinant that describes the full ensemble of molecules in a cavity, these symmetries can generate other (usually non-orthogonal) Hartree-Fock solutions with the same orbital energies~\cite{helgaker_molecular_2000}. These different solutions, which are called symmetry-broken solutions~\cite{li_symmetry_2009,kowalski_towards_1998,thom_hartreefock_2009}%[cite "Symmetry breaking in spin-restricted Hartree–Fock solutions: the case of the C2 molecule and the N+ 2 and F+ 2 cations", Xiangzhu Li and Josef Paldus, "Towards Complete Solutions to Systems of Nonlinear Equations of Many-Electron Theories" Karol Kowalski and Karol Jankowski, "Hartree–Fock solutions as a quasidiabatic basis for nonorthogonal configuration interaction" Alex J. W. Thom; Martin Head-Gordon] 
can correspond to different physical realizations of the ensemble. In the present case that is that there are different ways in which the electrons of the ensemble of molecules can react to nuclear/ionic and displacement-field changes. Similar to the non-orthogonal configuration-interaction ansatz \cite{thom_hartreefock_2009}
%[cite "Hartree–Fock solutions as a quasidiabatic basis for non-orthogonal configuration interaction" Alex J. W. Thom; Martin Head-Gordon] 
we collect all these different relevant Hartree-Fock solutions and assume that they share (at least approximately) most of the occupied (core) orbitals. We then choose one reference Hartree-Fock determinant $|\Phi_k^{N_o} \rangle$ and use the other highest-occupied and lowest-unoccupied molecular orbitals to generate an active space (which, for simplicity, we orthogonalize with respect to the reference Slater determinant) over which we perform a full diagonalization (see Fig. \ref{fig:occupation}). As a final assumption we consider %the quasi-dilute limit, i.e., the different molecules interact with each other
only electronic collective correlations from the long-range dipole-self-energy $\hat{W}_\perp$ term and we discard Coulombic contributions $\hat{W}_\parallel$. In this way, we focus on the inter-molecular correlation effects induced by the cavity. We finally note that the number of the different Hartree-Fock solutions, and with this the number of quasi-degenerate orbitals in the configuration-interaction ansatz, strongly depends on the nuclear/ionic and displacement-field state. It is here where synchronization (resonance condition) and symmetry considerations  become important. \cite{pang_role_2020, jayachandran_phenomenological_2025}

Our choice $k$ selects a single HF determinant from $\binom{N_o^{\rm tot}}{N_o}$ possible (degenerate) states with $N_o$-occupied orbitals in the $N_o^{\rm tot}$-fold degenerate subspace.
Then we perform a FCI expansion in this degenerate subspace. The virtual occupation number can be measured as follows $\hat{N}_v=\sum_{p \in virt}a_p^\dagger a_p $ with $\hat{N}_v|\Phi_{m_1...m_l}^{p_1...p_l}\rangle=l|\Phi_{m_1...m_l}^{p_1...p_l}\rangle$, for every $l$-fold excited Slater-determinant $|\Phi_{m_1...m_l}^{p_1...p_l}\rangle=a_{p_1}^\dagger...a_{p_l}^\dagger a_{m_l}...a_{m_1}|\Phi_k^{N_o} \rangle$ with $l_{\max}=\min(N_o,N_u)$ and $N_u=N_o^{\rm tot}-N_o$.
 
One would obtain  the exact transversal correlation energy in this subspace by calculating
%\begin{eqnarray}
 $   E_{\rm corr}=\bra{\Phi^{\rm FCI}}\hat{W}_\perp\ket{\Phi^{\rm FCI}}-E_k^{N_o}$
%\end{eqnarray}
with  subspace FCI wavefunction $\ket{\Phi^{\rm FCI}}=s_0|\Phi_k^{N_o} \rangle
  + \sum_{\boldsymbol{m}\in {\rm occ}}\!\sum_{\boldsymbol{p}\in {\rm unocc}} s_{m_1...m_l}^{p_1...p_l}\,|\Phi_{m_1...m_l}^{p_1...p_l}\rangle$, i.e., the excitations from the reference state remain restricted to the degenerate HF energy space with $\tilde{\epsilon}_o$.
However, generically, this is a highly non-trivial problem to solve. To obtain analytical physical insights, we proceed by assuming uniformity, i.e., equal probability weight per determinant $1/\binom{N_o^{\rm tot}}{N_o}$. The count of rank-l determinants is given by $\binom{N_o}{l}\binom{N_u}{l}$ and thus $w_l=\binom{N_o}{l}\binom{N_u}{l}/\binom{N_o^{\rm tot}}{N_o}$ yielding,
\begin{eqnarray}
    \langle \hat{N}_v\rangle_{\rm uniform}=\sum_{l=0}^{l_{\max}}l w_l=\frac{N_o N_u}{N_o^{\rm tot}}.\label{eq:nv_uniform}
\end{eqnarray}
and therefore we approximate,
\begin{eqnarray}
    E_{\rm corr}\approx\langle \hat{N}_v\rangle_{\rm uniform} E_{\rm CIS,corr}\label{eq:corr_approx}
\end{eqnarray}
where $E_{\rm CIS,corr}$ corresponds to the transversal electron correlations  derived from a configuration-interaction-singles (CIS) ansatz with fixed $N_o, N_u$~\cite{sidler_collectively-modified_2026}, 
\begin{equation}
|\Phi^\mathrm{CIS}_k\rangle
= s_0|\Phi_k \rangle
  + \sum_{m\in N_o}\!\sum_{p\in N_u} s_m^p\,|\Phi_m^p\rangle,
\end{equation}
which obeys $\langle\Phi^\mathrm{CIS}_k|\hat{N}_v|\Phi^\mathrm{CIS}_k\rangle=1$ by construction and thus justifies Eq. \eqref{eq:corr_approx}.
Orthogonality imposes the normalization condition $\langle\Phi^\mathrm{CIS}_k|\Phi^\mathrm{CIS}_ k\rangle=s_0^2+\sum_{m,p}(s_m^p)^2 =1 $.
As recently shown in Ref.~\cite{sidler_collectively-modified_2026}, the CIS transversal correlation energy problem can be mapped onto the spherical Sherrington-Kirkpatrick (SSK) model of a spin glass.  The resulting correlation energy minimization problem on the degenerate  subspace has the following form \cite{sidler_collectively-modified_2026}
\begin{equation}
E_{\rm CIS,corr} =- \sum_{i< j}^{N_{\rm deg}} J_{ij}\,\sigma_i\sigma_j,
  \qquad \sum_{i=1}^{N_{\rm deg}}\sigma_i^2 = 1%-s_0^2=\epsilon,
\label{eq:ECIS}
\end{equation}
where  $N_{\rm deg}=N_o N_u$ and $\sigma_i$ are normalized CIS amplitudes emergent from $s_m^p\mapsto \sigma_i$. They play the role of continuous spin variables that are normalized to one (i.e., the unknown weight of the reference state $s_0$ has been absorbed in random transversal interactions $J_{ij}$). Those emerge from two-electron integrals of the dipole-self energy, i.e., from $\hat{W}_\perp$, in combination with random molecular orientation (rotation) with respect to the linear cavity-polarization~\cite{sidler_collectively-modified_2026}. 
The spin glass problem in Eq. \eqref{eq:ECIS} has an analytic solution if the random-interaction are assumed to belong to the equivalence class of zero mean with extensive variance 
\begin{eqnarray}
     {\rm Var}(J_{ij})=\sigma^2 N_{\rm deg}, \label{eq:varjij}
 \end{eqnarray}
which emerge from
\begin{eqnarray}
    {\rm Var}(J_{ij}) = \lambda^4 e^4{\rm Var}(\bra{\phi_m}\hat{z}\ket{\phi_n}\bra{\phi_q}\hat{z}\ket{\phi_p}),\label{eq:sigma}
\end{eqnarray}
assuming a cavity polarization along the $z$-axis. The variable $\sigma^2$ quantifies the variance of random transversal interactions/fluctuations according to Eqs. \eqref{eq:varjij} and \eqref{eq:sigma}.
Consequently, the electron CIS correlation minimization problem $E_{\rm CIS,corr}$ is equivalent to the spherical Sherrington-Kirkpatrick (SSK) model of a spin glass.~\cite{sherrington_solvable_1975,baik_spherical_2021} Its analytic solution for  the  free energy in physicist's notation is given as $F_{\rm corr}=-k_B T \log(Z_{N_{\rm deg}})=N_{\rm deg} f$, with free energy per spin
%\begin{widetext}
\begin{eqnarray}
    f(T,\sigma)
    =&\begin{cases}
        -  \frac{\sigma^2}{4 k_B T}, & \sigma\leq k_B T \\
        -\sigma +\frac{k_B T}{2}\log{\big(\frac{\sigma }{k_B T}\big)}+\frac{3 k_B T}{4}, &\sigma> k_B T
    \end{cases}
    \label{eq:freeenergy}
\end{eqnarray}
%\end{widetext}
and critical temperature $T_c=\sigma/k_B$. Notice, when deriving Eq.~\eqref{eq:freeenergy}, the thermodynamic limit is implicitly assumed, i.e., $N_{\rm deg}\rightarrow\infty$. 

Eq. \eqref{eq:corr_approx} indicates that there are two distinct physical mechanism that can lead to the required extensive scaling of random interactions of the SSK model, i.e., ${\rm Var}(\langle \hat{N}_v\rangle_{\rm uniform}J_{ij})\sim N_{\rm deg}\rightarrow\infty$. First, the random interactions  scale extensively already on CIS level with  $E_{\rm CIS,corr}=N_{\rm deg} f$ and  $\langle\hat{N}_v\rangle_{\rm uniform}=1$, which  implies either an (almost) empty or (almost) fully occupied degenerate collective state. This case suggests an overall \textit{insulating behavior} in the thermodynamic limit. Second, the virtual occupation scales extensively because of half-filling with $E_{\rm CIS,corr}=f$ and  $\langle\hat{N}_v\rangle_{\rm uniform}=N_{\rm deg} $. %${\rm Var}(\langle \hat{N}_v\rangle_{\rm uniform}J_{ij})=\langle \hat{N}_v\rangle_{\rm uniform}^2{\rm Var}(J_{ij})\sim N_{\rm deg}\rightarrow\infty$ with ${\rm Var}(J_{ij})\sim 1/N_{\rm deg}$ because of Eq. \eqref{eq:nv_uniform}. \textcolor{red}{i.e., in the metallic case we relate $\langle \hat{N}_v\rangle_{\rm uniform}=N_{\rm deg}$ and thus $E_{\rm CIS,corr}=F_{\rm corr}/N_{\rm deg}$  Adjust Fo definition...mention thermodynamic limit in this considerations.} Notice, a vanishing variance of the random $J_{ij}$-couplings is a priori reasonable, when assuming the usual collective re-normalization of the light-matter coupling $\lambda^2 \propto 1/N$. 
The resulting fractional filling in the thermodynamic limit suggests a (cavity-induced) \textit{metallic behavior} of the transversal electron correlations, i.e., a high mobility of the $N_o$ occupied orbitals within the degenerate subspace. Both mechanisms (insulator, metal) allow to enter the SSK regime with an extensive transversal correlation free energy $F_{\rm corr}$, however, with different microscopic origin. 

%This means all considerations in the following assume that the occupied molecular ensemble is also considered in the thermodynamic limit, i.e., $N\sim N_{\rm deg}$.  In our cavity picture, the SSK free-energy corresponds to the long-range transversal correlation energy of the strongly couple molecular ensemble. Therefore, when assuming Eq.~\eqref{eq:eps}, the resulting total ensemble (free)-energy minimization problem in Eq.~\eqref{eq:ensemble} becomes
%\begin{eqnarray}
%    E = \min_{\boldsymbol{n}}\big[E_{\rm EHF}(\boldsymbol{n})+F(T,\sigma,\epsilon(\boldsymbol{n}))\big],\label{eq:min}
%\end{eqnarray}
%Notice at this point, that for simplicity, temperature and entropic considerations are limited to the free-energy of the (long-range) transversal electron correlations and discarded for $E_{\rm EHF}$, dominated by the Coulomb interaction.

In a next step, the important question arises, if and under which conditions, the transversal electron correlation energy can indeed play a chemically relevant role. %and under which conditions $\epsilon>0$ is energetically favorable. Furthermore, what are the conditions that the collectively correlated electrons undergo a spin glass phase transition? 
%or this purpose, we minimize
%Before we continue, one notices that the solving the standard literature CIS approach $E_{\rm tot}=\min_{s_0}\big[(1-\epsilon)E_{0}+F(T,\sigma,\epsilon)\big]$
%suggest $\epsilon_{\min}\in \{0,1\}$ and thus $\epsilon_{\min}=0$  for any physically reasonable case, i.e., no collective electron correlations emerges and $E_{\rm tot}=E_{0}$. In this trivial case, collective electron-correlations would only occur, if  the HF energy of the entire molecular ensemble  obeys $E_0>F(T,\sigma,\epsilon)$, which implies any molecular structure is lost and the HF reference state has lost any physical connection to this extremely correlated state.
%%%%%%%%%%%%%%%%%%%%
%\begin{eqnarray}
%    E_{\boldsymbol{n}} = \sum_{i=1}^N n_i\tilde{\epsilon_i}=
%\sum_{i=1}^N \big[n_i\epsilon_i +\frac{1}{2}\sum_{j=1}^N n_i n_j (2J_{ij}-K_{ij})\big], \label{eq:EHF}
%\end{eqnarray}
%with $n_i \in [0,2]$, and
%%
%
To answer this question, we will primarily focus on the metallic case, because it allows interesting qualitative analytical insights, whereas the insulating case requires the solution and study of a specific microscopic system which obeys ${\rm Var}(J_{ij})\sim N_{\rm deg}\rightarrow\infty$ to provide further qualitative insights. In that sense, if one assumes this condition to be met, transversal correlation effects always emerge by construction. Therefore, the detailed conditions for the emergence of transversal correlations in an insulating phase  goes beyond the scope of this work and will be the focus of a future publication. 

\begin{figure}
    \centering
    \includegraphics[width=1\linewidth]{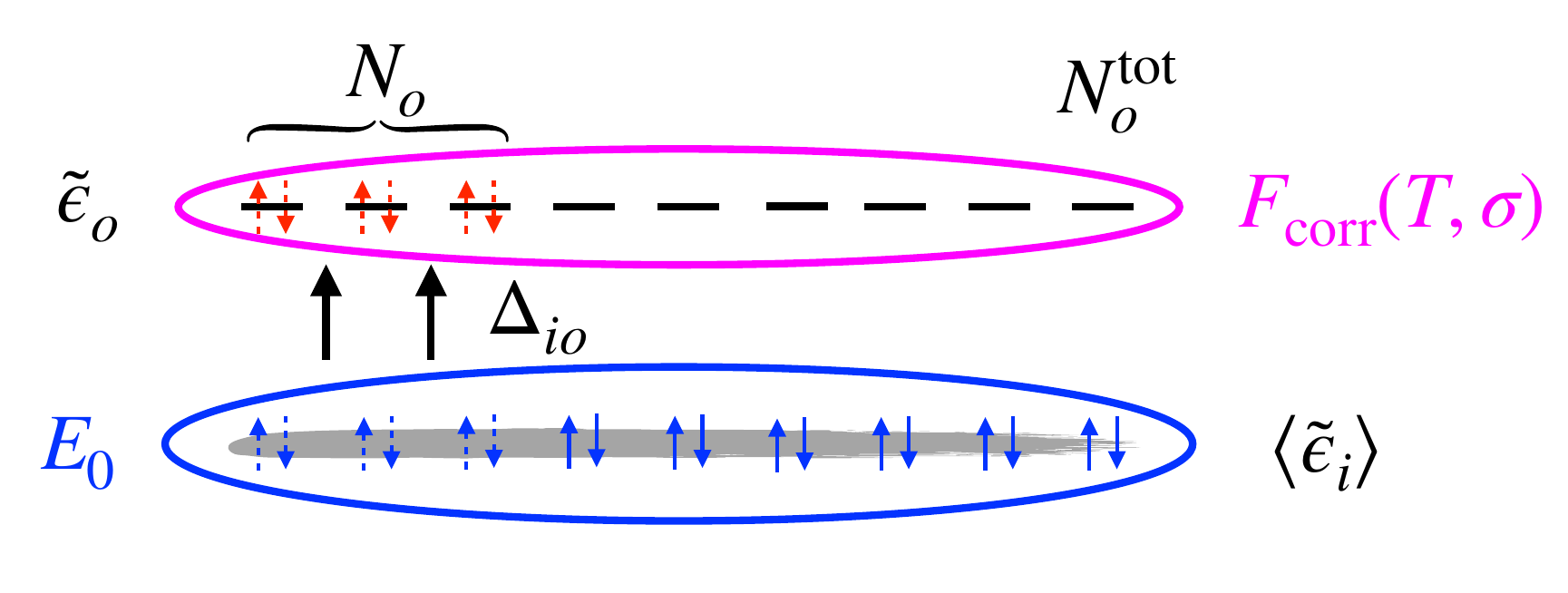}
    \caption{Simplified collective 2-level model to qualitatively investigate the formation of metallic transversal electron correlation phases under collective SC in the thermodynamic limit. Mean HF energies are labeled $\langle\tilde{\epsilon}_i\rangle $ which jointly accumulate to the  Hartree-Fock energy $E_0$ (blue). The cavity-induced $N_o^{\rm tot}$-fold degenerate space  is occupied $N_o$-times with excitation energy  $\Delta_{io}=\tilde{\epsilon}_o-\langle\tilde{\epsilon}_i\rangle>0$. Depending on the chemical ensemble, exciting $N_o$ orbitals can lower the transversal correlation free-energy $F_{\rm corr}$, and thus becoming energetically favorable. This suggest the formation of metallic transversal electron correlation phases.  }
    \label{fig:metal}
\end{figure}

\begin{figure}
    \centering
    \includegraphics[width=1\linewidth]{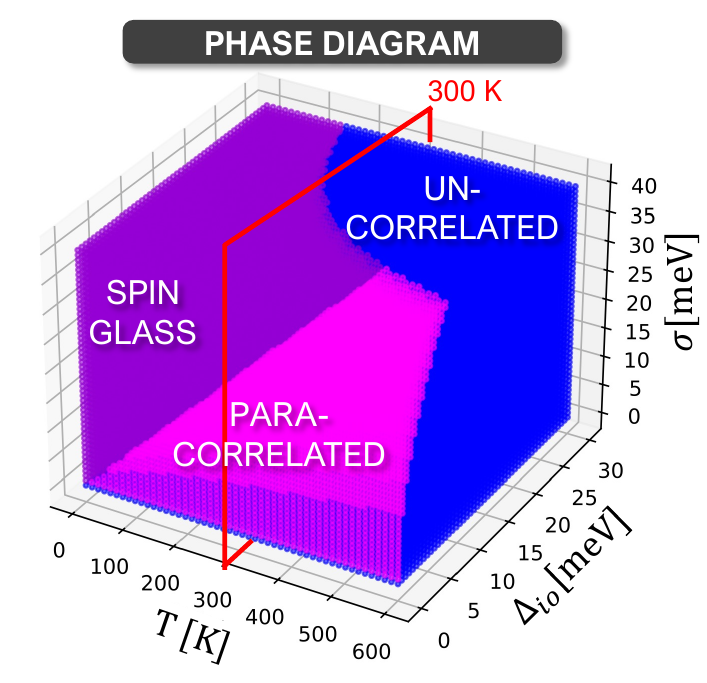}
    \caption{ Collective correlation phases in an ensemble of molecules subject to collective strong coupling in an optical cavity. The cavity-mediated synchronization of the electron fluctuations, give rise to random transversal electron-electron interactions (fluctuations) described by $\sigma$ (see Eq. \eqref{eq:sigma}). The HF energy,  to excite one orbital to the collectively correlated state, is given by $\Delta_{io}$. If the entropy of this highly degenerate  state dominates, fractional filling occurs and thus transversal electron correlations emerge, a metallic, para-correlated phase (paramagnetic in the spin glass model) is entered (pink). 
 At sufficient low temperatures, the SSK model exhibits a spin glass phase (purple). In this regime the time-evolution of the collectively correlated electrons should be governed by aging effects with a break down of fluctuation dissipation relations. Notice that the existence of collectively correlated phases seem plausible even at room temperature, because our model does not rely on macroscopic quantum coherence,\cite{mondal_macroscopic_2026} i.e., it just modifies the free space Coulomb electron dispersion forces, which are essential ingredients for many chemical phenomena at room temperature.  }
    \label{fig:phase}
\end{figure}

To yield qualitative insights into the cavity-induced metallic phase, however, we can investigate the conditions under which it is formed. For this purpose, we have the following simplified situation in mind (see. Fig. \ref{fig:metal}), where a single reference state shall possess non-degenerate doubly-occupied energy levels with average energy $\langle \tilde{\epsilon}_i\rangle$. Suppose we excite now $N_o$ orbitals $i$ with mean HF energy $\langle \tilde{\epsilon}_i\rangle$ into a collectively correlated (degenerate) state $\tilde{\epsilon}_o$. This excitation from a non-degenerate $\langle \tilde{\epsilon}_i\rangle$ to a degenerate state $\tilde{\epsilon}_o$ costs the average HF energy $\Delta_{io}=|\langle \tilde{\epsilon}_i\rangle-\tilde{\epsilon}_o|>0$.  If  increasing $N_o$ reduces the transversal correlation free energy sufficiently, however, a finite occupation of the degenerate state  $N_o>0$ can become favorable and thus the strongly coupled molecular ensemble may enter the above-defined metallic phase and thus the HF reference state is no longer an accurate description of our molecular ensemble. Remark: A holistic free energy picture for all electrons would in principle involve  all occupied and unoccupied HF orbitals.  However, for bare molecular ensembles (only longitudinal Coulomb interactions) at ambient conditions,  temperature and thus entropic effects are usually chemically irrelevant. Therefore, above restriction to the thermodynamic picture to the transversal correlations seems appropriate.\cite{wang_self-consistent-field_2022,graf_capturing_2026}     
Next, to investigate the entropic effect of the collective transversal interactions in our simplified model, we minimize the total energy,
\begin{eqnarray}
    E=E_0  -N_o\langle \tilde{\epsilon}_i\rangle +N_o \tilde{\epsilon}_o+\langle\hat{N}_v\rangle_{\rm uniform}f
\end{eqnarray}
with respect to $N_o$ and $N_o^{\rm tot}$, where the HF reference energy is given by $E_0=\bra{\Phi_k}\hat{H}_e\ket{\Phi_k}$. The minimum is found at:
\begin{eqnarray}
    N_o^*&=&\frac{N_o^{\rm tot}}{2 f}\big(f+\Delta_{io}\big)\label{eq:No}\\
    N_o^{\rm tot*}&\rightarrow&\infty,\label{eq:Ntot}
\end{eqnarray}
yield the following interpretation: First of all, Eq. \eqref{eq:Ntot} confirms that a collective degeneracy at $\tilde{\epsilon}_o$ is energetically (entropically) favorable. The emergence / existence of such a degeneracy $N_o^{\rm tot*}(\boldsymbol{R}, q_\beta)$ depends on the chemical ensemble of interest. Consequently, in practice the collective electron minimization problem implicitly depends on the external nuclear and displacement field coordinates. We expect that this is where the vibrational coupling (resonance, symmetry and synchronization) enter the holistic picture, i.e., what are the local chemical / implications of  $N_o^{\rm tot*}(\boldsymbol{R}, q_\beta)\rightarrow\infty$. We keep this aspect for future work and focus here instead on the interpretation of the electronic occupation in Eq. \eqref{eq:No}.
For $f< -\Delta_{io}$ a collectively correlated phase forms with (metallic) non-vanishing fractional filling $N_o^*/N_o^{\rm tot*}>0$. In other words, if the free energy reduction $f$ per single virtual excitation can compensate for the Coulomb-dominated HF interaction energy $\Delta E=N_o\Delta_{io}$ of the excited orbital, a novel collectively correlated phase of matter is formed with the following generic phase diagram in the thermodynamic limit (see Fig. \ref{fig:phase}) 
%\begin{widetext}
\begin{eqnarray}
%\boxed{
    E = \begin{cases}
        E_0& {\rm if}\ f(T,\sigma)\geq -\Delta_{io}\\
        E_0+ \Delta E+F_{\rm corr}(T,\sigma), & {\rm otherwise}%{\rm if} F(T,\sigma,1)< -E^{k,C}_{\rm Ion}(1)
    \end{cases}.%}
    \label{eq:correlation_spinglass}
\end{eqnarray}
%\end{widetext}
%and $E_{0}=\min_{\boldsymbol{n}}E_{\rm EHF}(\boldsymbol{n})$.
%Therefore, the collective scaling of the cluster-size removes the previous no go temperature criterion.
Eq. \eqref{eq:correlation_spinglass} reveals three distinct phases of electron correlations for molecular ensembles under collective strong coupling that depend on microscopic (local) properties of the molecular ensemble described by $\Delta_{io}$.  The \textit{uncorrelated} phase (only Coulomb correlations)  is found if the free energy of the transverse correlation cannot excite / stabilize a sufficient amount of electrons to reach a non-vanishing filling factor, and thus collective transverse correlations are absent. If the light-matter coupling is strong enough, such that $f< -\Delta_{io}$ and Eq. \ref{eq:Ntot} holds, our molecular ensemble may enter   the paramagnetic (\textit{paracorrelated}) solution of the SSK model. Remark: To determine if a specific molecular ensemble can enter the paracorrelated phase is a highly non-trivial question, since $\sigma$ and thus $F_{\rm corr}$ depend on the molecular ensemble under investigation. In particular, $\sigma$ depends parametrically on $(\boldsymbol{R},q_\beta)$. %In more detail, knowing the experimentally observed normal-incidence and resonance conditions and the general experimental struggle to induce chemical changes by collective SC, we expect their parametric impact on $\sigma$ to be relevant for the formation of collective correlations. Disentangling these aspect will require considerable future research efforts. 
Finally, from the SSK model one sees that a \textit{spin glass correlated} phase can emerge, at sufficiently low temperature $T$ or for strong collective fluctuations $\sigma$, i.e., for $\sigma>k_B T$. The para- to spin glass phase transition  is continuous for the free-energy, it may still affect changes the probability distributions of the spin variables, and introduces aging dynamics.

%%%%%%%%

In conclusion, we have demonstrated that collective transverse electron correlations can emerge  from collective degeneracies under strong coupling conditions. Two distinct mechanisms have been identified to cause transverse electron correlation effects. First, the insulating case, where the collectively degenerate state can be considered empty or fully occupied in the thermodynamic limit. Still its fluctuating electronic charges are  sufficiently strong to enter the  thermodynamic regime of the SSK spin glass model. Second, the metallic case, where a collectively degenerate state is fractionally occupied by an entropic reduction of the associated  transversal correlation free energy. The metallic correlations require less pronounced electronic fluctuations to enter the thermodynamic regime of the SSK model. 
Consequently, the collectively correlated SSK regime suggests the formation of two distinct phases of matter: a paracorrelated phase and a spin-glass correlated phase. %The formation of these highly degenerate phases is thermodynamically plausible at room temperature, provided that random dipole fluctuations, subsumed in  $\sigma$, are sufficiently strong — a condition governed by both the collective light-matter coupling strength and the degeneracy of the molecular ensemble. 
Formation and stabilization of these phases at room temperature  seems plausible if the transverse free energy gain is sufficiently large to populate and structure (induce collective degeneracy) the participating orbitals, a process that admits a natural interpretation as entropic orbital excitation into a metallic, collective electron correlation phase, and which constitutes a cavity-induced electron localization mechanism.
On a fundamental level, the indistinguishability of electrons — embodied in the Pauli exclusion principle — is essential to all these phenomena. The collectively correlated phases discussed here cannot emerge for distinguishable particles or in the dilute gas limit, two approximations that are widely invoked in models of collective strong coupling but that suppress precisely the quantum statistical effects responsible for the physics uncovered here. This distinction carries important implications for how collective strong coupling should be modeled in realistic molecular systems.
Our findings connect naturally to a growing body of experimental evidence. The relevance of cavity-modified dispersion interactions and phase transitions for chemistry has been directly observed experimentally,\cite{patrahau_direct_2024,sandeep_cluster_2026} and the emergence of collective electron correlation phases provides a compelling microscopic picture for recent Rayleigh-scattering experiments probing the electronic structure under collective vibrational strong coupling.\cite{sandeep_cluster_2026} Those experiments report a temperature-dependent first-order phase transition in the Rayleigh scattering amplitude — changing by up to two orders of magnitude beyond a critical coupling strength — with the onset of the Rabi splitting appearing to be independent of the observed phase transition. Furthermore, our observed formation of a fractionally occupied, collectively degenerate state, would suggest a rather metallic behavior of collective transverse electron correlations, which should appear in conductivity measurements. Indeed, experiments have shown an enhancement of the electrical conductivity by 6 orders of magnitude for intrinsically nonconducting  polymers such as polystyrene, deuterated polystyrene, and poly(benzyl methacrylate).\cite{kumar_extraordinary_2024} 
Looking ahead, several physically important questions remain open. For example, under which chemical conditions transversal electron correlation are dominated by metallic (fractionally occupied degeneracy) features vs.  insulating behavior. Moreover, it will be essential to clarify how cavity resonance conditions and molecular synchronization govern the applicability of the spherical Sherrington–Kirkpatrick model, and to establish precise connections between our predictions and the correlated methods developed for few-molecule strong coupling. More broadly, one may ask whether the collective localization mechanism uncovered here has relevance beyond electromagnetic cavities — extending, perhaps, to other forms of confined or structured environments in chemistry and materials science. We believe the answer is yes, and that the entropically driven collective electron correlations identified in this work represent a general and previously overlooked mechanism with implications reaching well beyond the cavity setting.

\begin{acknowledgments}
 This work was made possible through the support of the European Research Council (ERC-2024-SYG-101167294, UnMySt), the Cluster of Excellence Advanced Imaging of Matter (AIM), Grupos Consolidados (IT1249-19) and SFB925.   We acknowledge support from the Max Planck-New York City Center for Non-Equilibrium Quantum Phenomena. The Flatiron Institute is a division of the Simons Foundation.
\end{acknowledgments}
\bibliography{references}

@article{graf_capturing_2026,
	title = {Capturing electron correlation at mean-field cost: {Assessment} of i-{DMFT} and the underlying correlation conjecture},
	volume = {165},
	issn = {0021-9606, 1089-7690},
	shorttitle = {Capturing electron correlation at mean-field cost},
	url = {http://arxiv.org/abs/2604.19804},
	doi = {10.1063/5.0338943},
	number = {3},
	urldate = {2026-07-24},
	journal = {The Journal of Chemical Physics},
	author = {Graf, Paul G. and Matz, Florian and Ding, Lexin and Liebert, Julia and Penz, Markus and Schilling, Christian},
	month = jul,
	year = {2026},
	note = {arXiv:2604.19804 [physics.chem-ph]},
	pages = {034123},
}

@article{wang_self-consistent-field_2022,
	title = {Self-{Consistent}-{Field} {Method} for {Correlated} {Many}-{Electron} {Systems} with an {Entropic} {Cumulant} {Energy}},
	volume = {128},
	issn = {0031-9007, 1079-7114},
	url = {https://link.aps.org/doi/10.1103/PhysRevLett.128.013001},
	doi = {10.1103/PhysRevLett.128.013001},
	language = {en},
	number = {1},
	urldate = {2026-07-24},
	journal = {Physical Review Letters},
	author = {Wang, Jian and Baerends, Evert Jan},
	month = jan,
	year = {2022},
	pages = {013001},
}

@article{thom_hartreefock_2009,
	title = {Hartree–{Fock} solutions as a quasidiabatic basis for nonorthogonal configuration interaction},
	volume = {131},
	issn = {0021-9606, 1089-7690},
	url = {https://pubs.aip.org/jcp/article/131/12/124113/895960/Hartree-Fock-solutions-as-a-quasidiabatic-basis},
	doi = {10.1063/1.3236841},
	language = {en},
	number = {12},
	urldate = {2026-07-23},
	journal = {The Journal of Chemical Physics},
	author = {Thom, Alex J. W. and Head-Gordon, Martin},
	month = sep,
	year = {2009},
	pages = {124113},
}

@article{kowalski_towards_1998,
	title = {Towards {Complete} {Solutions} to {Systems} of {Nonlinear} {Equations} of {Many}-{Electron} {Theories}},
	volume = {81},
	copyright = {http://link.aps.org/licenses/aps-default-license},
	issn = {0031-9007, 1079-7114},
	url = {https://link.aps.org/doi/10.1103/PhysRevLett.81.1195},
	doi = {10.1103/PhysRevLett.81.1195},
	language = {en},
	number = {6},
	urldate = {2026-07-23},
	journal = {Physical Review Letters},
	author = {Kowalski, Karol and Jankowski, Karol},
	month = aug,
	year = {1998},
	pages = {1195--1198},
}

@article{li_symmetry_2009,
	title = {Symmetry breaking in spin-restricted {Hartree}–{Fock} solutions: the case of the {C2} molecule and the {N2}+ and {F2}+ cations},
	volume = {11},
	issn = {1463-9076, 1463-9084},
	shorttitle = {Symmetry breaking in spin-restricted {Hartree}–{Fock} solutions},
	url = {https://pubs.rsc.org/cp/article/11/26/5281-5289/98400},
	doi = {10.1039/b900184k},
	language = {en},
	number = {26},
	urldate = {2026-07-23},
	journal = {Physical Chemistry Chemical Physics},
	author = {Li, Xiangzhu and Paldus, Josef},
	year = {2009},
	pages = {5281},
}

@article{olsen_determinant_1988,
	title = {Determinant based configuration interaction algorithms for complete and restricted configuration interaction spaces},
	volume = {89},
	issn = {0021-9606, 1089-7690},
	url = {https://pubs.aip.org/jcp/article/89/4/2185/95039/Determinant-based-configuration-interaction},
	doi = {10.1063/1.455063},
	language = {en},
	number = {4},
	urldate = {2026-07-23},
	journal = {The Journal of Chemical Physics},
	author = {Olsen, Jeppe and Roos, Björn O. and Jo/rgensen, Poul and Jensen, Hans Jo/rgen Aa.},
	month = aug,
	year = {1988},
	pages = {2185--2192},
}

@article{mondal_macroscopic_2026,
	title = {A macroscopic condensation theory for vibrational strong coupling effects},
	issn = {2041-1723},
	url = {https://www.nature.com/articles/s41467-026-75222-2},
	doi = {10.1038/s41467-026-75222-2},
	language = {en},
	urldate = {2026-07-14},
	journal = {Nature Communications},
	author = {Mondal, M. Elious and Montillo Vega, Sebastian and Huo, Pengfei},
	month = jul,
	year = {2026},
}

@article{sandeep_cluster_2026,
	title = {Cluster {Formation} and {Phase} {Transitions} {Induced} by {Vibrational} {Strong} {Coupling}},
	volume = {138},
	issn = {0044-8249, 1521-3757},
	url = {https://onlinelibrary.wiley.com/doi/10.1002/ange.202516917},
	doi = {10.1002/ange.202516917},
	language = {en},
	number = {1},
	urldate = {2026-05-01},
	journal = {Angewandte Chemie},
	author = {Sandeep, K. and Swaminathan, S. and Jayachandran, A. and Nagarajan, K. and Gautier, J. and Kushida, S. and Chervy, T. and Vergauwe, R.M.A. and Thomas, A. and Ebbesen, T.W.},
	month = jan,
	year = {2026},
	pages = {e16917},
}

@article{sidler_collectively-modified_2026,
	title = {Collectively-{Modified} {Intermolecular} {Electron} {Correlations}: {The} {Connection} of {Polaritonic} {Chemistry} and {Spin} {Glass} {Physics}: {Focus} {Review}},
	volume = {126},
	copyright = {https://creativecommons.org/licenses/by/4.0/},
	issn = {0009-2665, 1520-6890},
	shorttitle = {Collectively-{Modified} {Intermolecular} {Electron} {Correlations}},
	url = {https://pubs.acs.org/doi/10.1021/acs.chemrev.4c00711},
	doi = {10.1021/acs.chemrev.4c00711},
	language = {en},
	number = {1},
	urldate = {2026-05-01},
	journal = {Chemical Reviews},
	author = {Sidler, Dominik and Ruggenthaler, Michael and Rubio, Angel},
	month = jan,
	year = {2026},
	pages = {4--27},
}

@article{jayachandran_phenomenological_2025,
	title = {A {Phenomenological} {Symmetry} {Rule} for {Chemical} {Reactivity} {Under} {Vibrational} {Strong} {Coupling}},
	volume = {137},
	issn = {0044-8249, 1521-3757},
	url = {https://onlinelibrary.wiley.com/doi/10.1002/ange.202503915},
	doi = {10.1002/ange.202503915},
	language = {de},
	number = {35},
	urldate = {2025-10-20},
	journal = {Angewandte Chemie},
	author = {Jayachandran, A. and Patrahau, B. and Ricca, J. G. and Mahato, M. K. and Pang, Y. and Nagarajan, K. and Thomas, A. and Genet, C. and Ebbesen, T. W.},
	month = aug,
	year = {2025},
	pages = {e202503915},
}

@article{pang_role_2020,
	title = {On the {Role} of {Symmetry} in {Vibrational} {Strong} {Coupling}: {The} {Case} of {Charge}‐{Transfer} {Complexation}},
	volume = {59},
	issn = {1433-7851, 1521-3773},
	shorttitle = {On the {Role} of {Symmetry} in {Vibrational} {Strong} {Coupling}},
	url = {https://onlinelibrary.wiley.com/doi/10.1002/anie.202002527},
	doi = {10.1002/anie.202002527},
	language = {en},
	number = {26},
	urldate = {2025-05-22},
	journal = {Angewandte Chemie International Edition},
	author = {Pang, Yantao and Thomas, Anoop and Nagarajan, Kalaivanan and Vergauwe, Robrecht M. A. and Joseph, Kripa and Patrahau, Bianca and Wang, Kuidong and Genet, Cyriaque and Ebbesen, Thomas W.},
	month = jun,
	year = {2020},
	pages = {10436--10440},
}

@article{kumar_extraordinary_2024,
	title = {Extraordinary {Electrical} {Conductance} through {Amorphous} {Nonconducting} {Polymers} under {Vibrational} {Strong} {Coupling}},
	volume = {146},
	copyright = {https://doi.org/10.15223/policy-029},
	issn = {0002-7863, 1520-5126},
	url = {https://pubs.acs.org/doi/10.1021/jacs.4c03016},
	doi = {10.1021/jacs.4c03016},
	language = {en},
	number = {28},
	urldate = {2025-05-25},
	journal = {Journal of the American Chemical Society},
	author = {Kumar, Sunil and Biswas, Subha and Rashid, Umar and Mony, Kavya S. and Chandrasekharan, Gokul and Mattiotti, Francesco and Vergauwe, Robrecht M. A. and Hagenmuller, David and Kaliginedi, Veerabhadrarao and Thomas, Anoop},
	month = jul,
	year = {2024},
	pages = {18999--19008},
}

@book{szabo_modern_2012,
	address = {Mineola, New York},
	title = {Modern quantum chemistry: introduction to advanced electronic structure theory},
	isbn = {978-0-486-69186-2 978-0-486-13459-8},
	shorttitle = {Modern quantum chemistry},
	language = {eng},
	publisher = {Dover Publications, Inc},
	author = {Szabo, Attila and Ostlund, Neil S.},
	year = {2012},
}

@article{baik_spherical_2021,
	title = {Spherical {Spin} {Glass} {Model} with {External} {Field}},
	volume = {183},
	issn = {0022-4715, 1572-9613},
	url = {https://link.springer.com/10.1007/s10955-021-02757-7},
	doi = {10.1007/s10955-021-02757-7},
	language = {en},
	number = {2},
	urldate = {2025-02-04},
	journal = {Journal of Statistical Physics},
	author = {Baik, Jinho and Collins-Woodfin, Elizabeth and Le Doussal, Pierre and Wu, Hao},
	month = may,
	year = {2021},
	pages = {31},
}

@book{helgaker_molecular_2000,
	address = {Chichester New York},
	title = {Molecular electronic-structure theory},
	isbn = {978-1-119-01957-2 978-1-119-01956-5},
	language = {eng},
	publisher = {Wiley},
	author = {Helgaker, Trygve and Olsen, Jeppe and Jørgensen, Poul},
	year = {2000},
	doi = {10.1002/9781119019572},
}

@article{marsh_entanglement_2024,
	title = {Entanglement and {Replica} {Symmetry} {Breaking} in a {Driven}-{Dissipative} {Quantum} {Spin} {Glass}},
	volume = {14},
	issn = {2160-3308},
	url = {https://link.aps.org/doi/10.1103/PhysRevX.14.011026},
	doi = {10.1103/PhysRevX.14.011026},
	language = {en},
	number = {1},
	urldate = {2024-09-19},
	journal = {Physical Review X},
	author = {Marsh, Brendan P. and Kroeze, Ronen M. and Ganguli, Surya and Gopalakrishnan, Sarang and Keeling, Jonathan and Lev, Benjamin L.},
	month = feb,
	year = {2024},
	pages = {011026},
}

@article{guo_emergent_2019,
	title = {Emergent and broken symmetries of atomic self-organization arising from {Gouy} phase shifts in multimode cavity {QED}},
	volume = {99},
	issn = {2469-9926, 2469-9934},
	url = {https://link.aps.org/doi/10.1103/PhysRevA.99.053818},
	doi = {10.1103/PhysRevA.99.053818},
	language = {en},
	number = {5},
	urldate = {2024-09-19},
	journal = {Physical Review A},
	author = {Guo, Yudan and Vaidya, Varun D. and Kroeze, Ronen M. and Lunney, Rhiannon A. and Lev, Benjamin L. and Keeling, Jonathan},
	month = may,
	year = {2019},
	pages = {053818},
}

@article{guo_sign-changing_2019,
	title = {Sign-{Changing} {Photon}-{Mediated} {Atom} {Interactions} in {Multimode} {Cavity} {Quantum} {Electrodynamics}},
	volume = {122},
	issn = {0031-9007, 1079-7114},
	url = {https://link.aps.org/doi/10.1103/PhysRevLett.122.193601},
	doi = {10.1103/PhysRevLett.122.193601},
	language = {en},
	number = {19},
	urldate = {2024-09-19},
	journal = {Physical Review Letters},
	author = {Guo, Yudan and Kroeze, Ronen M. and Vaidya, Varun D. and Keeling, Jonathan and Lev, Benjamin L.},
	month = may,
	year = {2019},
	pages = {193601},
}

@article{gopalakrishnan_exploring_2012,
	title = {Exploring models of associative memory via cavity quantum electrodynamics},
	volume = {92},
	issn = {1478-6435, 1478-6443},
	url = {http://www.tandfonline.com/doi/abs/10.1080/14786435.2011.637980},
	doi = {10.1080/14786435.2011.637980},
	language = {en},
	number = {1-3},
	urldate = {2024-09-19},
	journal = {Philosophical Magazine},
	author = {Gopalakrishnan, Sarang and Lev, Benjamin L. and Goldbart, Paul M.},
	month = jan,
	year = {2012},
	pages = {353--361},
}

@article{gopalakrishnan_frustration_2011,
	title = {Frustration and {Glassiness} in {Spin} {Models} with {Cavity}-{Mediated} {Interactions}},
	volume = {107},
	copyright = {http://link.aps.org/licenses/aps-default-license},
	issn = {0031-9007, 1079-7114},
	url = {https://link.aps.org/doi/10.1103/PhysRevLett.107.277201},
	doi = {10.1103/PhysRevLett.107.277201},
	language = {en},
	number = {27},
	urldate = {2024-09-19},
	journal = {Physical Review Letters},
	author = {Gopalakrishnan, Sarang and Lev, Benjamin L. and Goldbart, Paul M.},
	month = dec,
	year = {2011},
	pages = {277201},
}

@article{ebbesen_hybrid_2016,
	title = {Hybrid {Light}–{Matter} {States} in a {Molecular} and {Material} {Science} {Perspective}},
	volume = {49},
	issn = {0001-4842, 1520-4898},
	url = {https://pubs.acs.org/doi/10.1021/acs.accounts.6b00295},
	doi = {10.1021/acs.accounts.6b00295},
	language = {en},
	number = {11},
	urldate = {2024-09-09},
	journal = {Accounts of Chemical Research},
	author = {Ebbesen, Thomas W.},
	month = nov,
	year = {2016},
	pages = {2403--2412},
}

@article{ruggenthaler_understanding_2023,
	title = {Understanding {Polaritonic} {Chemistry} from {Ab} {Initio} {Quantum} {Electrodynamics}},
	volume = {123},
	issn = {0009-2665, 1520-6890},
	url = {https://pubs.acs.org/doi/10.1021/acs.chemrev.2c00788},
	doi = {10.1021/acs.chemrev.2c00788},
	language = {en},
	number = {19},
	urldate = {2023-11-12},
	journal = {Chemical Reviews},
	author = {Ruggenthaler, Michael and Sidler, Dominik and Rubio, Angel},
	month = oct,
	year = {2023},
	pages = {11191--11229},
}

@article{sidler_perspective_2022,
	title = {A perspective on ab initio modeling of polaritonic chemistry: {The} role of non-equilibrium effects and quantum collectivity},
	volume = {156},
	issn = {0021-9606},
	shorttitle = {A perspective on ab initio modeling of polaritonic chemistry},
	url = {https://doi.org/10.1063/5.0094956},
	doi = {10.1063/5.0094956},
	number = {23},
	urldate = {2023-09-22},
	journal = {The Journal of Chemical Physics},
	author = {Sidler, Dominik and Ruggenthaler, Michael and Schäfer, Christian and Ronca, Enrico and Rubio, Angel},
	month = jun,
	year = {2022},
	pages = {230901},
}

@article{xiang_molecular_2024,
	title = {Molecular {Polaritons} for {Chemistry}, {Photonics} and {Quantum} {Technologies}},
	volume = {124},
	copyright = {https://creativecommons.org/licenses/by/4.0/},
	issn = {0009-2665, 1520-6890},
	url = {https://pubs.acs.org/doi/10.1021/acs.chemrev.3c00662},
	doi = {10.1021/acs.chemrev.3c00662},
	language = {en},
	number = {5},
	urldate = {2024-09-05},
	journal = {Chemical Reviews},
	author = {Xiang, Bo and Xiong, Wei},
	month = mar,
	year = {2024},
	pages = {2512--2552},
}

@article{mandal_theoretical_2023,
	title = {Theoretical {Advances} in {Polariton} {Chemistry} and {Molecular} {Cavity} {Quantum} {Electrodynamics}},
	volume = {123},
	copyright = {https://creativecommons.org/licenses/by/4.0/},
	issn = {0009-2665, 1520-6890},
	url = {https://pubs.acs.org/doi/10.1021/acs.chemrev.2c00855},
	doi = {10.1021/acs.chemrev.2c00855},
	language = {en},
	number = {16},
	urldate = {2024-09-05},
	journal = {Chemical Reviews},
	author = {Mandal, Arkajit and Taylor, Michael A.D. and Weight, Braden M. and Koessler, Eric R. and Li, Xinyang and Huo, Pengfei},
	month = aug,
	year = {2023},
	pages = {9786--9879},
}

@article{simpkins_control_2023,
	title = {Control, {Modulation}, and {Analytical} {Descriptions} of {Vibrational} {Strong} {Coupling}},
	volume = {123},
	copyright = {https://doi.org/10.15223/policy-001},
	issn = {0009-2665, 1520-6890},
	url = {https://pubs.acs.org/doi/10.1021/acs.chemrev.2c00774},
	doi = {10.1021/acs.chemrev.2c00774},
	language = {en},
	number = {8},
	urldate = {2024-09-05},
	journal = {Chemical Reviews},
	author = {Simpkins, Blake S. and Dunkelberger, Adam D. and Vurgaftman, Igor},
	month = apr,
	year = {2023},
	pages = {5020--5048},
}

@article{hirai_molecular_2023,
	title = {Molecular {Chemistry} in {Cavity} {Strong} {Coupling}},
	volume = {123},
	copyright = {https://doi.org/10.15223/policy-029},
	issn = {0009-2665, 1520-6890},
	url = {https://pubs.acs.org/doi/10.1021/acs.chemrev.2c00748},
	doi = {10.1021/acs.chemrev.2c00748},
	language = {en},
	number = {13},
	urldate = {2024-09-05},
	journal = {Chemical Reviews},
	author = {Hirai, Kenji and Hutchison, James A. and Uji-i, Hiroshi},
	month = jul,
	year = {2023},
	pages = {8099--8126},
}

@article{bhuyan_rise_2023,
	title = {The {Rise} and {Current} {Status} of {Polaritonic} {Photochemistry} and {Photophysics}},
	volume = {123},
	copyright = {https://creativecommons.org/licenses/by/4.0/},
	issn = {0009-2665, 1520-6890},
	url = {https://pubs.acs.org/doi/10.1021/acs.chemrev.2c00895},
	doi = {10.1021/acs.chemrev.2c00895},
	language = {en},
	number = {18},
	urldate = {2024-09-05},
	journal = {Chemical Reviews},
	author = {Bhuyan, Rahul and Mony, Jürgen and Kotov, Oleg and Castellanos, Gabriel W. and Gómez Rivas, Jaime and Shegai, Timur O. and Börjesson, Karl},
	month = sep,
	year = {2023},
	pages = {10877--10919},
}

@article{fregoni_theoretical_2022,
	title = {Theoretical {Challenges} in {Polaritonic} {Chemistry}},
	volume = {9},
	copyright = {https://creativecommons.org/licenses/by/4.0/},
	issn = {2330-4022, 2330-4022},
	url = {https://pubs.acs.org/doi/10.1021/acsphotonics.1c01749},
	doi = {10.1021/acsphotonics.1c01749},
	language = {en},
	number = {4},
	urldate = {2024-09-05},
	journal = {ACS Photonics},
	author = {Fregoni, Jacopo and Garcia-Vidal, Francisco J. and Feist, Johannes},
	month = apr,
	year = {2022},
	pages = {1096--1107},
}

@article{ruggenthaler_quantum-electrodynamical_2018,
	title = {From a quantum-electrodynamical light–matter description to novel spectroscopies},
	volume = {2},
	issn = {2397-3358},
	url = {https://www.nature.com/articles/s41570-018-0118},
	doi = {10.1038/s41570-018-0118},
	language = {en},
	number = {3},
	urldate = {2024-09-05},
	journal = {Nature Reviews Chemistry},
	author = {Ruggenthaler, Michael and Tancogne-Dejean, Nicolas and Flick, Johannes and Appel, Heiko and Rubio, Angel},
	month = mar,
	year = {2018},
	pages = {0118},
}

@article{ebbesen_introduction_2023,
	title = {Introduction: {Polaritonic} {Chemistry}},
	volume = {123},
	copyright = {https://doi.org/10.15223/policy-001},
	issn = {0009-2665, 1520-6890},
	shorttitle = {Introduction},
	url = {https://pubs.acs.org/doi/10.1021/acs.chemrev.3c00637},
	doi = {10.1021/acs.chemrev.3c00637},
	language = {en},
	number = {21},
	urldate = {2024-09-05},
	journal = {Chemical Reviews},
	author = {Ebbesen, Thomas W. and Rubio, Angel and Scholes, Gregory D.},
	month = nov,
	year = {2023},
	pages = {12037--12038},
}

@article{sherrington_solvable_1975,
	title = {Solvable {Model} of a {Spin}-{Glass}},
	volume = {35},
	copyright = {http://link.aps.org/licenses/aps-default-license},
	issn = {0031-9007},
	url = {https://link.aps.org/doi/10.1103/PhysRevLett.35.1792},
	doi = {10.1103/PhysRevLett.35.1792},
	language = {en},
	number = {26},
	urldate = {2024-05-02},
	journal = {Physical Review Letters},
	author = {Sherrington, David and Kirkpatrick, Scott},
	month = dec,
	year = {1975},
	pages = {1792--1796},
}

@article{patrahau_direct_2024,
	title = {Direct {Observation} of {Polaritonic} {Chemistry} by {Nuclear} {Magnetic} {Resonance} {Spectroscopy}},
	issn = {1433-7851, 1521-3773},
	url = {https://onlinelibrary.wiley.com/doi/10.1002/anie.202401368},
	doi = {10.1002/anie.202401368},
	language = {en},
	urldate = {2024-05-02},
	journal = {Angewandte Chemie International Edition},
	author = {Patrahau, B. and Piejko, M. and Mayer, R. J. and Antheaume, C. and Sangchai, T. and Ragazzon, G. and Jayachandran, A. and Devaux, E. and Genet, C. and Moran, J. and Ebbesen, T. W.},
	month = may,
	year = {2024},
	pages = {e202401368},
}

 \end{document}